\documentclass[12pt]{article}
\usepackage{epsfig}

\def\beq{\begin{equation}}
\def\eeq#1{\label{#1}\end{equation}}
\def\eeqn{\end{equation}}

\def\beqa{\begin{eqnarray}}
\def\eeqa#1{\label{#1}\end{eqnarray}}
\def\eeqan{\end{eqnarray}}

\let\bar=\overbar

\def\Dslash{\not{\hbox{\kern-4pt $D$}}}
\def\dslash{\not{\hbox{\kern-2pt $\del$}}}

\def\msb{{\bar{\ssstyle M \kern -1pt S}}}

\def\Title#1{\begin{center} {\Large {\bf #1} } \end{center}}

\begin{document}

\Title{Searches for long-lived charged particles with the ATLAS experiment}

\bigskip\bigskip


\begin{raggedright}  

{\it Christian Ohm\index{Christian, C.} on behalf of the ATLAS Collaboration\\
Fysikum, Stockholm University}
\bigskip\bigskip
\end{raggedright}

\section{Introduction}
Stable massive particles (SMPs) offer spectacular detector signatures entirely without physics backgrounds. 
This paper summarizes the results of two searches for SMPs~\cite{Aad:2011yf,Aad:2011hz} performed with the ATLAS experiment 
at the LHC using $34$-$37$~pb$^{-1}$ from 2010.

\section{Outline of the searches}

Lepton-like SMPs, such as long-lived sleptons, are expected to lose energy through EM processes and give rise to a detector signature similar to that of slow-moving muons. Hadron-like SMPs, such as $R$-hadrons, could exchange electric charge when penetrating the detector material and therefore possibly be dominantly neutral in the inner tracker or the muon spectrometer~\cite{deBoer:2007ii,Farrar:2010ps,Mackeprang:2009ad}.
Therefore, two strategies were pursued. 

The first search~\cite{Aad:2011yf} targets hadron-like SMPs uses only the inner detector (ID) and calorimetry in ATLAS, without relying on the muon spectrometer (MS). In events triggered by a $E_{\mathrm{T}}^{\mathrm{miss}} > 40$~GeV signature, slow-particle candidates are sought as tracks with $p_{\mathrm{T}} > 50$~GeV in $|\eta| < 1.7$. The speed of the candidate is then estimated in two independent ways. Through the Bethe-Bloch relation $\beta\gamma$ is extracted from the ionization energy loss measurement in the Pixel detector, and $\beta$ is estimated from time-of-flight measurements in the Tile calorimeter. By combining the track momentum and speed measurements, two mass estimates are obtained through \mbox{$m = p/\beta\gamma$}. The signal regions are defined in the resulting two-dimensional mass plane, and a data-driven method is used to estimate the yields from background processes.

The second search~\cite{Aad:2011hz} relies on a signal in the MS. 
A muon trigger was used to collect the experimental data, and two selections were defined to target lepton-like and hadron-like SMPs separately. For the lepton-like SMPs, two candidates with $p_{\mathrm{T}} > 40$~GeV are required in the events, each with a combined track in both the ID and MS. For the hadron-like SMP selection only one candidate with $p_{\mathrm{T}} > 60$~GeV is required per event, and in the absence of combined tracks standalone MS tracks are also used, in order to be sensitive to $R$-hadrons that are neutral in the ID and charged in the MS. One single $\beta$ measurement is then formed using information from both the monitored drift tube precision chambers and the fast resistive plate trigger chambers in the muon spectrometer, as well as the Tile calorimeter where available. Combined with the track momentum measurement, a mass estimate can be calculated for each candidate. A  data-driven background estimate was also used for this search.



%
%

\section{Results}

\begin{figure}[htb]
\begin{center}
\includegraphics[width=0.56\textwidth]{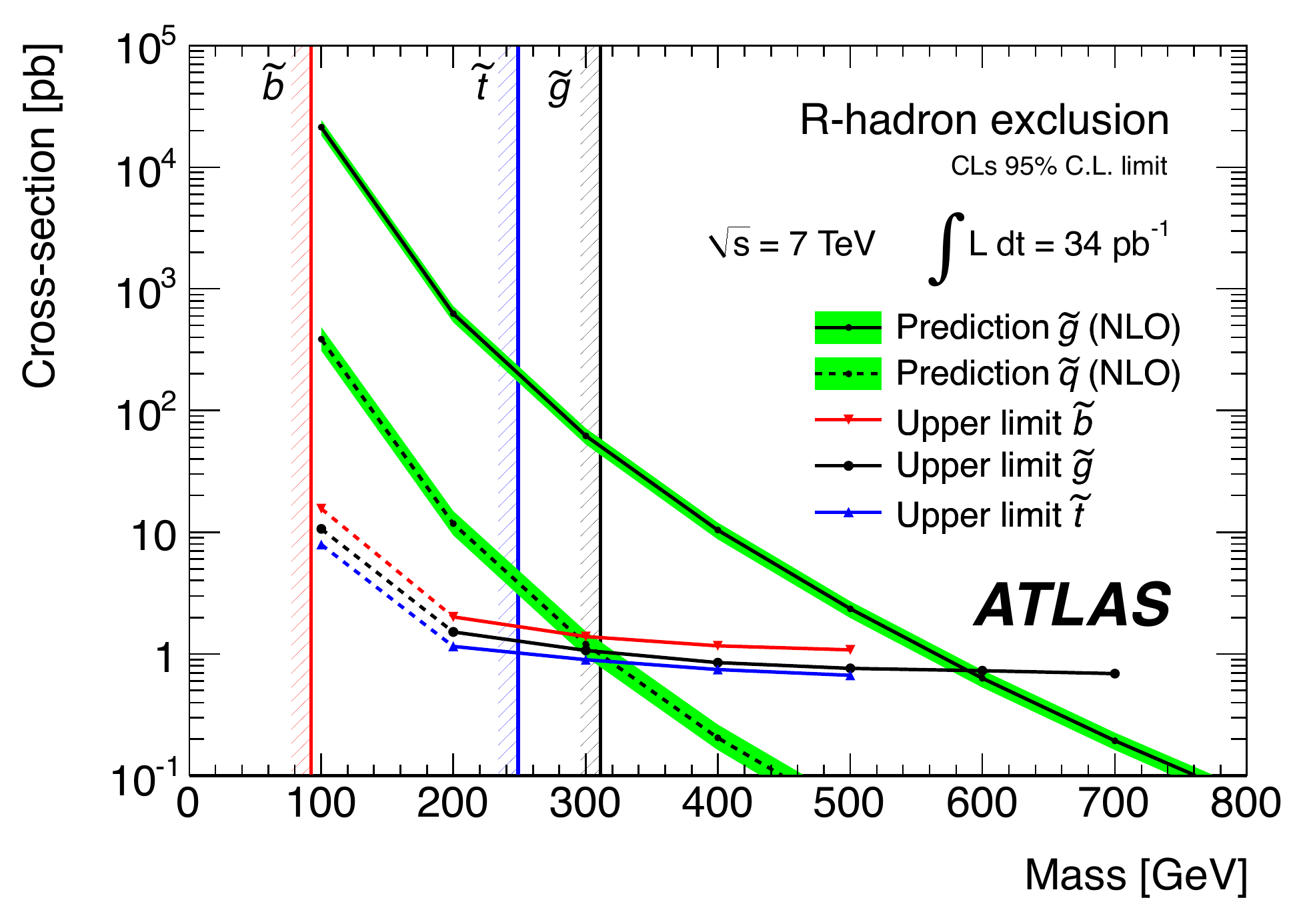}
\includegraphics[width=0.41\textwidth]{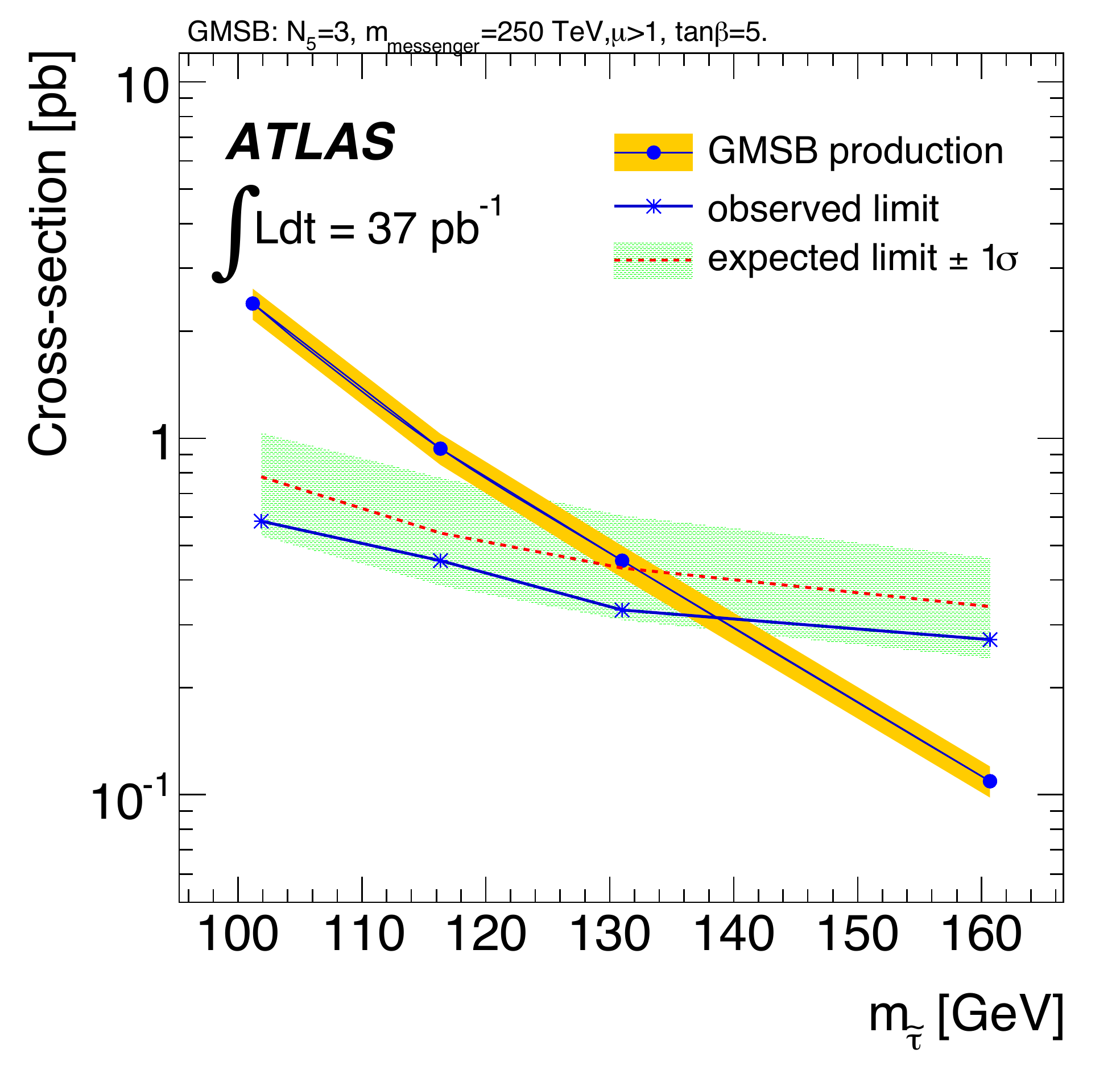}
\caption{Measured 95\% C.L. upper cross section limits for $R$-hadrons (left) and long-lived sleptons (right), along with predictions based on theoretical calculations.}
\label{fig:limits}
\end{center}
\end{figure}

In both searches the observations agree well with the low predicted yields due to instrumental effects in a background-only hypothesis, and 95\% C.L. cross section limits are calculated for the production cross section of each signal scenario. Figure~\ref{fig:limits} illustrates the extracted cross section limits along with theoretical predictions for production of $\tilde{g}$, $\tilde{t}$ and $\tilde{b}$ (left), and a model of gauge-mediated supersymmetry breaking (GMSB) featuring a stable $\tilde{\ell}$ (right). The intersections give the resulting mass constraints: $m_{\tilde{g}} > 586$~GeV, $m_{\tilde{t}} > 309$~GeV, $m_{\tilde{b}} > 284$~GeV and $m_{\tilde{\tau}} > 136$~GeV.

%
%
%
%
%
 

\begin{thebibliography}{99}


\bibitem{Fairbairn:2006gg}
  M.~Fairbairn {\it et al},
  Phys.\ Rept.\  {\bf 438}, 1 (2007)
  [arXiv:hep-ph/0611040].


\bibitem{Aad:2011yf}
  G.~Aad {\it et al.},  
  Phys.\ Lett.\  B {\bf 701}, 1 (2011)
  [arXiv:1103.1984 [hep-ex]].

\bibitem{Aad:2011hz}
  G.~Aad {\it et al.}, 
  Phys.\ Lett.\  B {\bf 703}, 428 (2011)
  [arXiv:1106.4495 [hep-ex]].

\bibitem{deBoer:2007ii}
 Y.~R.~de Boer {\it et al},
 J.\ Phys.\ G {\bf 35}, 075009 (2008)
 [arXiv:0710.3930 [hep-ph]].

\bibitem{Farrar:2010ps}
 G.~R.~Farrar {\it et al},
 JHEP {\bf 1102}, 018 (2011)
 [arXiv:1011.2964 [hep-ph]].

\bibitem{Mackeprang:2009ad}
 R.~Mackeprang and D.~Milstead,
 Eur.\ Phys.\ J.\  C {\bf 66}, 493 (2010)
 [arXiv:0908.1868 [hep-ph]].

\end{thebibliography}
\end{document}